\documentclass[%
reprint,
 amsmath,amssymb,
 aps,
showpacs,showkeys,]{revtex4-1}

\usepackage{graphicx}
\usepackage{dcolumn}
\usepackage{bm}

\begin{document}

\title{An approximation of the $S$ matrix for solving the Marchenko equation}

\author{N. A. Khokhlov}
\email{nikolakhokhlov@yandex.ru; nakhokhlov@gmail.com}
\affiliation{%
Peter the Great St. Petersburg Polytechnic University, St. Petersburg, Russia.
}%
\date{\today}%

\begin{abstract}
I present a new approximation of the $S$-matrix dependence on momentum $q$, formulated as a sum of a rational function and a truncated Sinc series. This approach enables pointwise determination of the $S$ matrix with specified resolution, capturing essential features such as resonance behavior with high accuracy. The resulting approximation provides a separable kernel for the Marchenko equation (fixed-$l$ inversion), reducing it to a system of linear equations for the expansion coefficients of the output kernel. Numerical results demonstrate good convergence of this method, applicable to both unitary and non-unitary $S$ matrices. Convergence is further validated through comparisons with an exactly solvable square-well potential model. The method is applied to analyze $S_{31}$ $\pi N$ scattering data.
\end{abstract}

\maketitle

\section{\label{sec:intro}INTRODUCTION}
In nuclear physics, a major challenge is the extraction of the interparticle interaction potential from scattering data. 
The inherent ill-posedness of this problem complicates its solution. Various methods have been employed to address this issue, including fitting the parameters of phenomenological potentials. While this approach is straightforward when data is sparse or of limited accuracy, more realistic cases  require precise methods for solving the quantum scattering inverse problem (IP) in order to accurately describe the substantial amount of experimental data. The development of such precise and unambiguous methods remains a significant challenge in the field \cite{Sparenberg1997,Sparenberg2004,Kukulin2004,Pupasov2011,Mack2012}. 
The scattering data, presented in the form of partial wave analysis (PWA) data with a fixed orbital momentum $l$, serve as natural inputs to the Marchenko, Krein, and Gelfand-Levitan theories \cite{Gelfand, Agranovich1963, Marchenko1977, Blazek1966, Krein1955, Levitan1984, Newton, Chadan}. 
These theories reformulate the inverse problem as the solution of a Fredholm integral equation of the second kind.
For a unique solution, these methods require knowledge of the binding energies, the corresponding bound-state normalization constants, and the $S$-matrix for all energies from zero to infinity. The latter may limit the applicability of such methods. However, relativistic two-particle potential models can be expressed in a nonrelativistic form \cite{Keister1991}, thereby extending the applicability of IP methods beyond nonrelativistic quantum mechanics.
In some elastic scattering processes, phase shifts behave as slowly varying smooth functions over a wide energy range, with resonant behavior occurring infrequently, even beyond the inelasticity threshold. This behavior suggests that the  potential model remains valid in these cases \cite{Funk2001}.

The Marchenko and Gelfand-Levitan methods were successfully utilized for extraction of $NN$  partial potentials from PWA data \cite{Geramb1994, Kohlhoff1994}. The analysis included data up to the inelastic threshold ($E_{\text{lab}} \approx 280$ MeV).
A similar approach, within the framework of Marchenko theory, was later employed to derive optical model $NN$ partial potentials describing elastic $NN$ scattering  data  up to 3 GeV \cite{Khokhlov2006, Khokhlov2007}. 
The Marchenko theory is applicable to nonunitary $S$-matrices that account for absorption \cite{Zakhariev1990}. It has been employed in the description of elastic neutron-deuteron ($nD$) scattering over a wide energy range, from zero to well above the threshold \cite{Papastylianos1990, Alt1994}. In this context, the Marchenko theory produces energy-independent complex partial potentials. 

In all such calculations until recently, partial $S$ matrices (spectral densities) were approximated using rational fraction expansions.
In this context, the input kernel of the integral equation is represented as finite separable series. For separable kernels, the Fredholm integral equation admits an analytical solution. Consequently, the partial potentials can be expressed in terms of Riccati-Hankel functions, leading to Bargmann-type potentials.
The advancement of IP methods in nuclear physics has been significantly limited by the conventional dependence on expressing the scattering matrix (spectral densities) through rational fraction expansions. 
Recently, a new method was proposed for solving the Marchenko equation by decomposing its input kernel into a separable series of isosceles triangular-pulse functions \cite{MyAlg2,MyAlg3}.Although this approach is more general and can, in principle, reproduce any $S$ matrix and converge, its accuracy is limited by the need to extract the potential via numerical differentiation of the output kernel.

In this work, a novel approximation for the momentum dependence of the $S$ matrix is proposed. The approximation consists of the sum of a rational function and a truncated Sinc series, allowing for pointwise determination of the $S$ matrix with a specified resolution. The proposed method results in a separable kernel for the Marchenko equation, transforming the integral equation into a system of linear equations for the expansion coefficients of the output kernel. This approach facilitates the direct determination of the potential from the Marchenko equation and an equation derived via its analytical differentiation, thereby circumventing the need for numerical differentiation of the output kernel.

\section{$S$-matrix parametrization for Marchenko inversion}
The radial Schrödinger equation for angular momentum $l$,
\begin{equation}
	\label{f1}
	\left(\frac{d^{2}}{d r^{2}}-\frac{l(l+1)}{r^{2}}-V(r)+q^{2}\right) \psi(r, q)=0,
\end{equation}
is linked with the Marchenko integral equation  \cite{Agranovich1963, Marchenko1977}
\begin{equation}
	\label{f3}	F(x, y)+L(x, y)+\int_{x}^{+\infty} L(x, t) F(t, y) d t=0.
\end{equation}
The input kernel is defined as 
\begin{multline}
	F(x, y)=\frac{1}{2 \pi} \int_{-\infty}^{+\infty} h_{l}^{+}(q x)[1-S(q)] h_{l}^{+}(q y) d q \\	
	+\sum_{j=1}^{n_{b}} h_{l}^{+}\left(\tilde{q}_{j} x\right) M_{j}^{2} h_{l}^{+}\left(\tilde{q}_{j} y\right)\\
	=\frac{1}{2 \pi} \int_{-\infty}^{+\infty} h_{l}^{+}(q x) Y(q) h_{l}^{+}(q y) d q
	\label{f4}	
\end{multline}
where $h_{l}^{+}(z)$ is the Riccati-Hankel function. 
The properties of the $S$ matrix and bound states required for the unique determination of the local potential are as follows
\begin{equation}
	\label{f2}
	\left\{S(q),(0<q<\infty), \tilde{q}_{j}, M_{j}, j=1, \ldots, n\right\}.
\end{equation}
where $S(q)$  is a scattering matrix dependence on the momentum $q$; 
$\tilde{q}_{j}^{2}=E_{j} \leq 0, E_{j}$  is $j$th bound state energy 
($-\imath \tilde{q}_{j} \geq 0$), $M_{j}$  is $j$th bound state asymptotic constant.

In case of non-unitary $S$ matrix, the absorbing partial $S$ matrix should be defined as  
\begin{equation}
	S(q)=\left\{\begin{array}{lll}
		S_{u}(q)+S_{n}(q) & \text { for } & q>0, \\
		S^{+}_{u}(-q)-S^{+}_{n}(-q) & \text { for } & q<0, 
	\end{array}\right.
	\label{OP_Smatrix}
\end{equation}
where superscript $+$ means hermitian conjugation. For $q>0$, the $S$ matrix is defined as 
\begin{equation}
	S_{u}(q) = e^{2\imath \delta(q)},\ \
	S_{n}(q) = -\sin^2(\rho(q))e^{2\imath \delta(q)}.
\end{equation}
Here,  $\delta(q)$ and $\rho(q)$ represent the phase shift and inelasticity parameter, respectively. These parameters may differ from those used in other works \cite{Arndt1989}. For a unitary $S$ matrix, $\rho(q)\equiv 0$ and $S_{n}(q)\equiv 0$. 

The potential function of Eq.~(\ref{f1}) is obtained from the output kernel 
\begin{equation}
	V(r)=-2 \frac{d L(r, r)}{d r} \label{f6}
\end{equation}

The solution of the IP Eqs.~\ref{f1}-\ref{f6} involves four distinct numerical steps: (1) performing PWA, (2) approximating the $S$ matrix, (3) solving the integral equation, and (4) differentiating the output kernel to compute the potential. Errors accumulate at each stage of the procedure, making an accurate assessment of the uncertainty in the final potential a nontrivial task.
The initial data are experimentally provided as confidence intervals for the values of the $S$ matrix at specific values of $q$. The challenge is to approximate the $S$ matrix by interpolating and extrapolating these experimentally obtained values in such a way that a separable kernel $F(x, y)$ is produced. 
To demonstrate his method, Marchenko utilized the rational representation of the $S$ matrix (rational function). Only this representation has been employed for the interpolation and extrapolation of $S$-matrix values until recently.

 In this case, the unitary $S$ matrix takes the form
\begin{equation} \label{Su_frac} 
	S_{0}(q)=e^{2 \imath \delta(q)} = \frac{f_2(q) + i f_1(q)}{f_2(q) - i f_1(q)}, 
\end{equation} 
where $f_1(q)$ and $f_2(q)$ are odd and even polynomials of $q$, respectively.
For this $S$ matrix, 
\begin{multline}
	\frac{1}{2 \pi} \int_{-\infty}^{+\infty} h_{l}^{+}(q x)[1-S_{0}(q)] h_{l}^{+}(q y) d q \\	
	= \sum\limits_{i = 1}^{n_{\mbox{\tiny{pos}}}} {b_i h_l^ + \left(
	{\beta _{i} x} \right)h_l^{+} \left( {\beta _{i} y}
	\right)}.	  
	\label{F_forS0}	
\end{multline}
Here, $\beta_i$ ($i = 1, \dots, n_{\mbox{\tiny{pos}}}$) represent all poles of $S_0$ with $\textrm{Im}[\beta_i] > 0$, and $b_i$ ($i = 1, \dots, n_{\mbox{\tiny{pos}}}$) are constants determined from the corresponding residues.

 Further, $[n/m]$ will denote the powers $n$ and $m$ of the polynomials $f_{1}$ and $f_{2}$, respectively. Since $\tan\delta = f_{1}/f_{2}$, we have $m = n + 1$. To interpolate the data $S(q_j), j = 1, 2, \dots, N$, a system of linear equations is derived from Eq.~\ref{Su_frac}, which allows for the determination of the coefficients of $f_1$ and $f_2$. In case of  data perturbed by random noise (errors) this  problem is generally ill-conditioned and may result in the emergence of spurious zero-pole pairs, known as Froissart doublets.
 For elastic $NN$ and $\pi N$ scattering, the available PWA data, particularly at high energies, are significantly affected by experimental errors. Furthermore, it is unclear whether the asymptotic behavior of the $S$ matrix is achieved at the energies studied. Accurately describing the available PWA data in this context necessitates the use of high-order polynomials in Eq.~(\ref{Su_frac}) \cite{Khokhlov2006, Khokhlov2007}. However, increasing the degree $[n/m]$ of the approximant to enhance accuracy may introduce substantial variations in the resulting potential. Thus, the rational approximant appears suboptimal for $S$-matrix approximation, as approximants of different orders that yield close $S$-matrix values at the PWA data points may diverge significantly between these points and at higher energies. Consequently, the convergence of methods that use Eq.~(\ref{Su_frac}) as the $S$-matrix approximation for the Marchenko equation is not guaranteed and remains uncertain.
The choice to approximate the $S$ matrix in our problem using Eq.~(\ref{Su_frac})  arises from its analytic structure, which yields a separable kernel for the Marchenko equation. This approximation is able to accurately capture the behavior of the phase shift $\delta(q)$ at both the small $q$ limit ($\delta \sim q^{l+1}$) and the large $q$ limit ($\delta \sim 1/q$). 
It is possible to retain the desirable properties of this approximation while overcoming its limitations. A successful approach was previously proposed \cite{MyAlg2,MyAlg3}; however, as noted, it encounters challenges in numerical implementation. From general principles, it is evident that an approximation in the form of a series with the required properties provides a solution. 
To do this, it is sufficient to approximate $Y(q)$ by a sum of the term of the form Eq.~(\ref{Su_frac})  and a series of functions with a finite support having a separable Fourier transform. As shown below, it is possible to construct functions with these properties for potentials with a finite range.
For  $l=0$,
\begin{equation}
	F(x, y)
	=\frac{1}{2 \pi} \int_{-\infty}^{+\infty} e^{q (x+y)} Y(q) d q.
	\label{f4L0}	
\end{equation}
Let $R$ be the interaction range, and  $2R\tau =\pi$.  There is  a set of functions with approximately finite support and having an approximately separable Fourier transform
\begin{eqnarray}
	\lambda_{k}(q)=\frac{\sin\left((k\tau-q)\pi/\tau\right)}{(k\tau-q)\pi/\tau}. \label{sin_basis_L0}
\end{eqnarray} 
The Fourier transform
\begin{multline}
	\frac{1}{2\pi}\int_{-\infty}^{\infty} \lambda_{k}(q) e^{\imath qz} d q \\
		=\frac{\tau}{2\pi} e^{\imath k\tau z}H(2R-|z|)|_{z=x+y} \\
			\approx \frac{\tau}{2\pi} e^{\imath k\tau (x+y)} H(R-|x|)H(R-|x|),
			\label{Fourier_l0}
		\end{multline}
	where $H(z)$ is the Heaviside step function.
	I approximate $	H(2R-|x+y|) \approx H(R-|x|)H(R-|x|)$ assuming that $F(x,y)\approx 0$ then $x,y >R$. These approximations become accurate for a potential with a range of $R$. Equation~(\ref{Fourier_l0})
can be generalized for $l>0$, 
\begin{multline}
	\frac{1}{2\pi} \int_{-\infty}^{+\infty} h_{l}^{+}(q x)
	\lambda_{k}(q)
	h_{l}^{+}(q y) d q	  \\
	=H(2R-|x+y|) 
	\frac{\tau}{2\pi}	
	h_{l}^{+}(k\tau x)h_{l}^{+}(k\tau y) \\
	\approx \frac{\tau}{2\pi}  H(R-|x|)H(R-|x|)h_{l}^{+}(k\tau x)h_{l}^{+}(k\tau y). 
	\label{Fourier_l1}
\end{multline}

I separate the asymptotic part  $S_{asymp}$ of the $S$ matrix as a rational function fit, 
\begin{equation}
	S(q)=S_{asymp}(q)+\Delta S(q).
	   \label{SmatrixN}
\end{equation}
 To account for absorption,  I employ a more flexible approximation in place of Eq.~(\ref{Su_frac}),
\begin{equation}
	S_{asymp}(q)=\frac{f_{2}(q)+\imath f_{1}(q)}{f_{2}(q)-\imath f_{1}(q)}+
	\frac{g_{1}(q)}{g_{2}(q)+\imath g_{1}(q)}.
	\label{SmatrixN2}
\end{equation}
Here $f_{1}(q)$ and $f_{2}(q)$ are odd and even polynomials of $q$, respectively, and $g_{1}(q)$ and $g_{2}(q)$ are odd and even polynomials of $q$, respectively.  
\begin{equation}
	\Delta S(q)=- \sum_{k=-N}^{N} s_{k} \lambda_{k}(q).
	\label{SmatrixN3}
\end{equation}
The coefficients are determined by 
\begin{equation}
	s_{k}  
	= -\left.\Delta S(q)\right|_{q=k\tau},\ \ 
		s_{0}\equiv 0.
	\label{ykm}
\end{equation}
For a unitary $S$ matrix, we have $g_{1}(q) \equiv 0$, and $s_{m,-k} = s_{m,k}^{*}$.
In this representation of the $S$ matrix, it is assumed that $S_{\text{asymp}}$ describes the behavior of the $S$ matrix at both small and large values of $q$, providing a rough approximation at intermediate values of $q$. Consequently, the degree of the rational approximation of $S$ should be minimized to avoid the appearance of spurious zero-pole pairs. 
The description of the $S$ matrix at intermediate values of $q$ is provided by a truncated Sinc series. 
		Then 
		\begin{equation}
			F(x, y) 
\approx \sum\limits_{j = 1}^n {b_j \Lambda_j \left( { x} \right)\Lambda_j \left( { y} \right)}
				, \label{FKrein2}
		\end{equation}
where
\begin{equation}
\Lambda_j \left( { x} \right) =
	H(R-|x|) { h}_{l}^{+} \left( {\beta _j x}\right). 
\end{equation}
	$\beta=\{(j\tau, j=-N..N),\beta_{1},..\beta_{n_{pos}},\tilde{q_{1}},...\tilde{q_{n_{b}}}\}$, $n=2N+n_{pos}+n_{b}$; 
	$\beta_i$ ($i = 1, \dots, n_{\mbox{\tiny{pos}}}$) represent all poles of $S_{asymp}$ with $\textrm{Im}[\beta_i] > 0$, and $b_i$ ($i = 1, \dots, n_{\mbox{\tiny{pos}}}$) are constants determined from the corresponding residues.
	I approximate  
	\begin{equation}
	 { h}_{l}^{+} \left( {\beta _j x}\right) \approx
				H(R-|x|) { h}_{l}^{+} \left( {\beta _j x}\right), 
				\end{equation}
		for transformation of rational fraction part  $S_{asymp}(q)$ of Eq.~(\ref{SmatrixN}) assuming that $F(x,y)\approx 0$ then $x,y >R$. 
I substitude 
\begin{equation}
	\label{f9n}	L(x, y) = \sum_{k=0}^{n}P_{k}(x)\Lambda_{k}(y),
\end{equation}
and Eq.~(\ref{FKrein2}) into Eq.~(\ref{f3}). Linear independence of the basis functions gives for $j=0..M$
	\begin{equation}
		\sum_{m=0}^{n}\left(\delta_{k\, m}+\zeta_{k\, m}(x) \right) P_{m}(x)=- b_{k}\Lambda_{k}(x),
		\label{f10n}	
	\end{equation}
	where $\delta_{j\, m}$ are the Kronecker symbols, and 
	\begin{equation}
		\zeta_{k\, m}(x)=b_{k}\int_{x}^{\infty} \Lambda_{k}(t) \Lambda_{m}(t) d t.
		\label{f10b}	
	\end{equation}

The solution of the linear system of Eqs.(\ref{f10n}) provides $P_{m}(x)$ at any point $x$. Since Eqs.~(\ref{f10n}) can be easily differentiated, $dP_{m}(x)/dx$ can be obtained directly from the corresponding system of linear equations. Consequently, the potential in Eq.~(\ref{f6}) is computed from the derivatives $dP_{m}(x)/dx$ and $d\Lambda_{m}(x)/dx$ without the need for numerical differentiation

\section{Results and Conclusions}

\begin{figure}[h]
	\centerline{\includegraphics[width=0.5\textwidth]{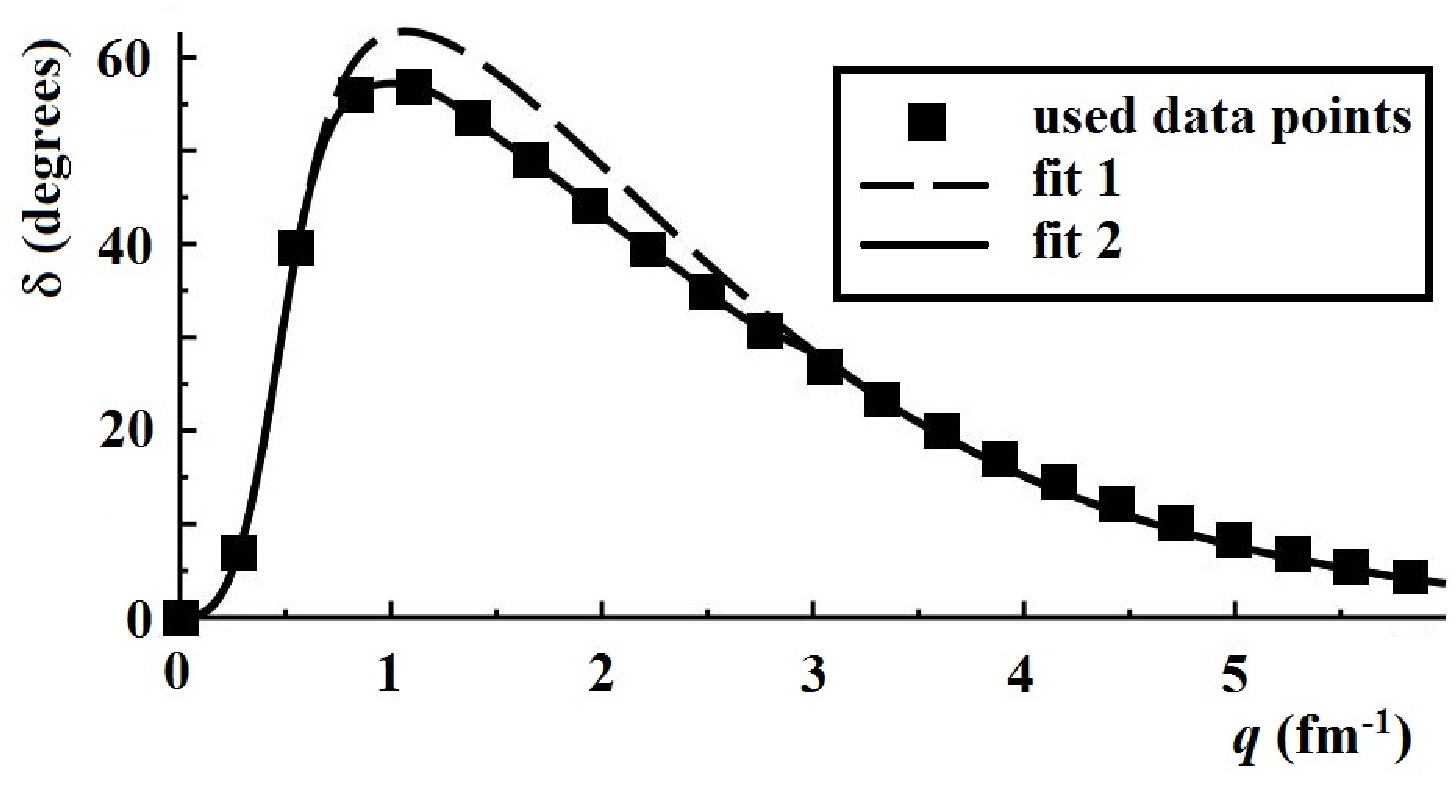}}
	\caption{\label{fig:testdata} Data (solid boxes) used to reconstruct a test potential $V(r)=V_{1}e^{-a_{0}r^2}+V_{2}e^{-a_{1}r}$,  where $V_{1}=20\ fm^{-2}= 829.8\ MeV$, $a_{1}=5.5\ fm^{-2}$, $V_{2}=-10\ fm^{-2}= 414.9\ MeV$, $a_{2}=1.5\ fm^{-1}$. Angular orbital momentum $l=1$. Units correspond to the $NN$ system. Phase curves for two fits of the $S$ matrix are shown. The dashed line corresponds to $S(q)=S_{asymp}(q)$ (rational function fit), while the solid line represents $S(q)=S_{asymp}(q)+\Delta S(q)$.
		\label{fig:example_data}
	}
\end{figure}

\begin{figure}[h]
	\centerline{\includegraphics[width=0.5\textwidth]{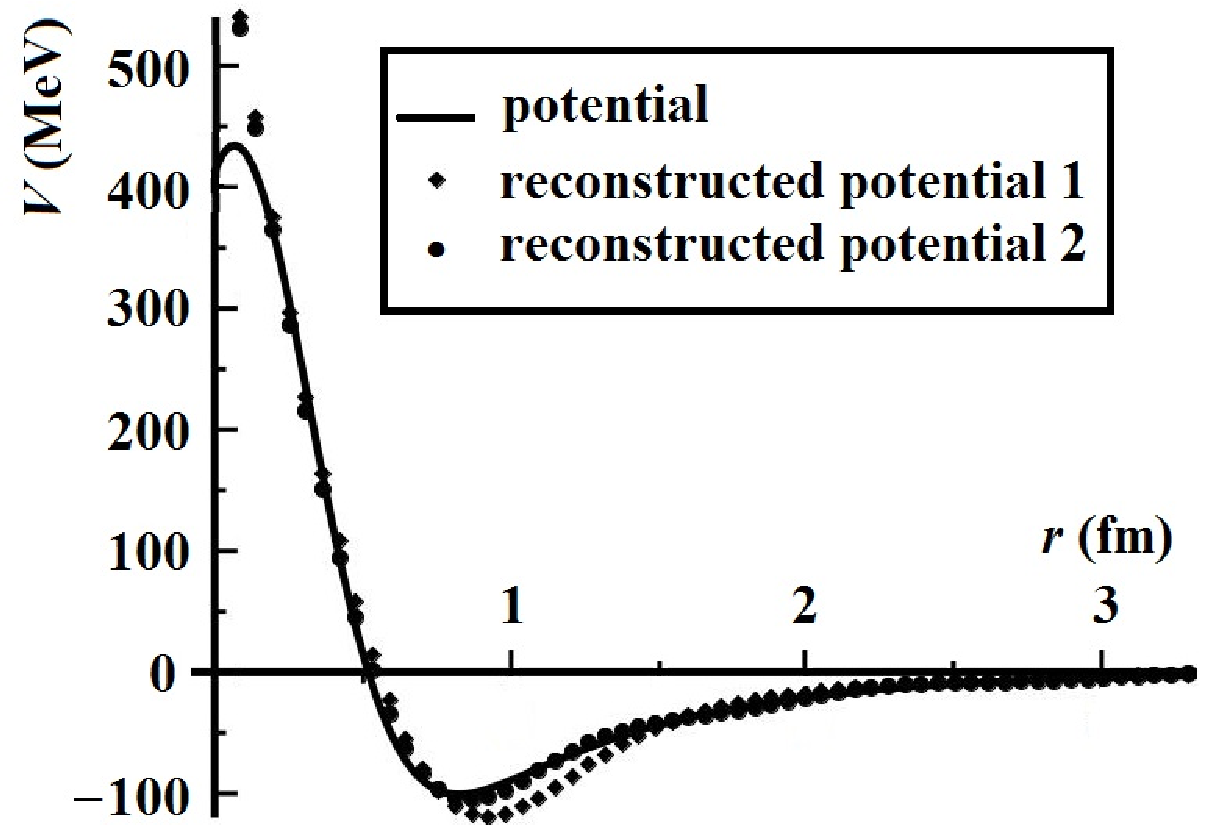}}
	\caption{\label{fig:example_pot} Results of inversion for $S$-matrix fits shown in Fig.~\ref{fig:testdata}. Angular orbital momentum $l=1$. Solid line is the initial potential  $V(r)=V_{1}e^{-a_{0}r^2}+V_{2}e^{-a_{1}r}$.  Solid diamonds represent the reconstruction from the rational function fit   $S(q)=S_{asymp}(q)$, while solid circles represent the reconstruction from $S(q)=S_{asymp}(q)+\Delta S(q)$.}
\end{figure}

\begin{figure}[h]
	\centerline{\includegraphics[width=0.5\textwidth]{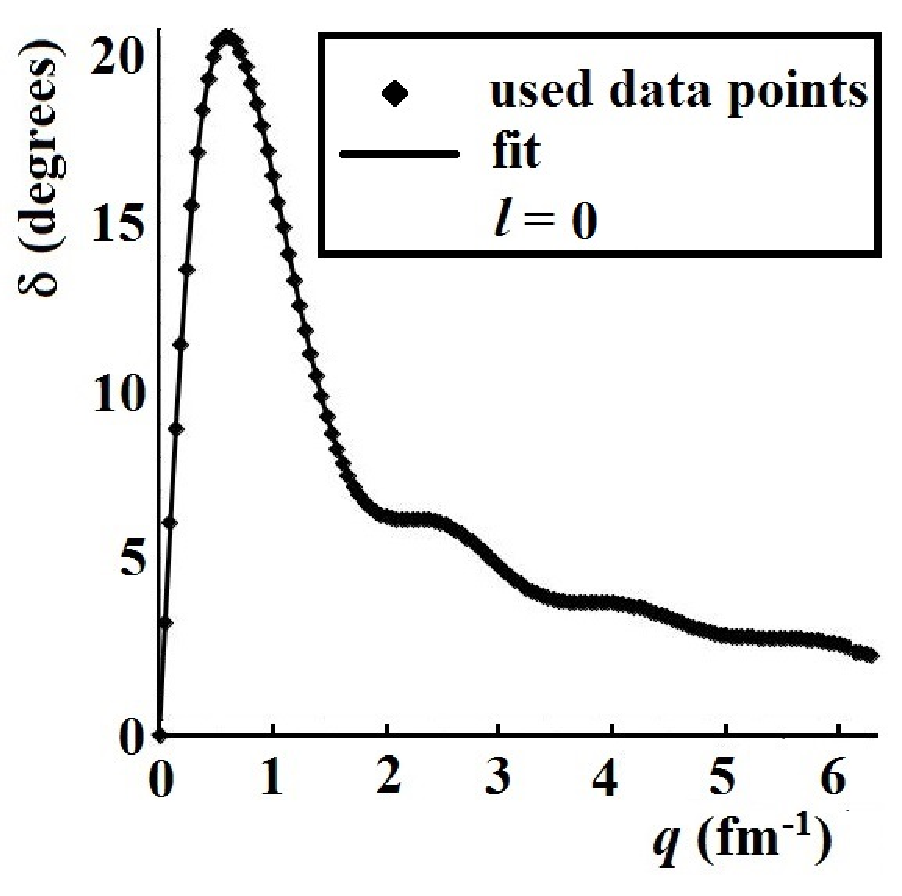}}
	\caption{\label{fig:example_data_pit} 
	Data (solid boxes) used to reconstruct a finite square well  potential $V(r)=-0.2569 \ fm^{-2}= 10\ MeV$, for $r<a=1.973\ fm$.  Angular orbital momentum $l=0$. A fit for $S$ matrix  $S(q)=S_{asymp}(q)+\Delta(q)$ is shown by a phase line.
	  }
\end{figure}

\begin{figure}[h]
	\centerline{\includegraphics[width=0.5\textwidth]{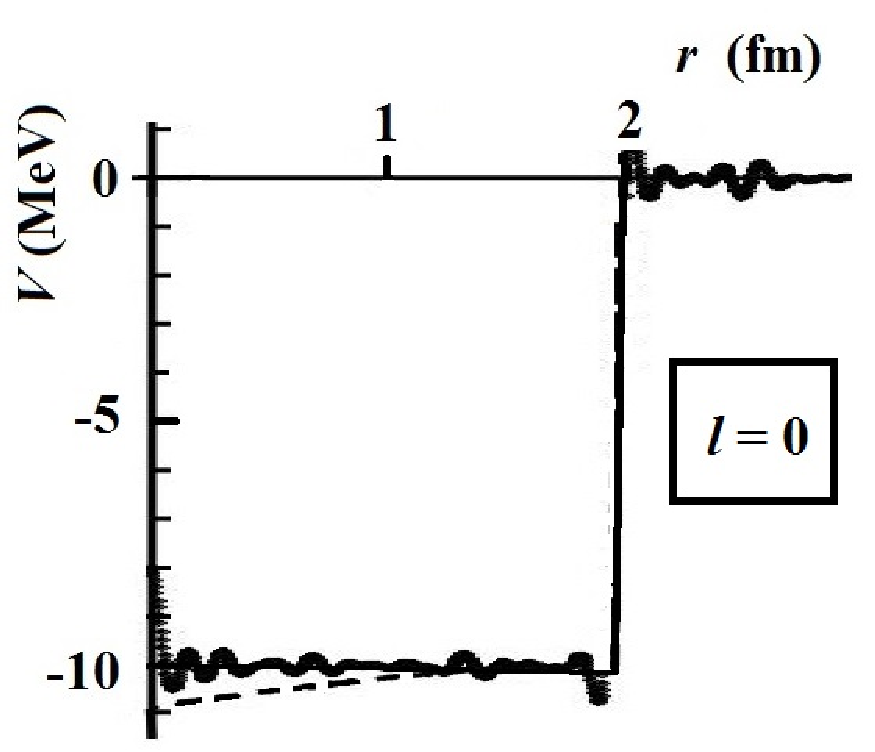}}
	\caption{\label{fig:example_pot_pit}
	Results of the inversion for the $S$-matrix fit shown in Fig.~\ref{fig:example_data_pit}. The dashed line corresponds to the method from Refs.\cite{MyAlg2,MyAlg3} (Method 1), while the solid thick line represents the method developed in this work (Method 2).}
\end{figure}

The developed approach was tested by reconstructing various potentials from their respective scattering data. Representative results for unitary $S$ matrices are shown in  Figs.~\ref{fig:example_data},~\ref{fig:example_pot},~\ref{fig:example_data_pit},~\ref{fig:example_pot_pit}. 
A rational approximation of the $S$ matrix can satisfactorily describe potentials with simple $\delta(q)$ dependencies, as demonstrated in Figs.~\ref{fig:example_data},~\ref{fig:example_pot}. For potentials with oscillatory phase dependence, fitting a rational approximation to the $S$ matrix presents significant challenges. The approach of Ref.\cite{MyAlg1}, and the method introduced in this work, effectively addresses these issues, as demonstrated in  Figs.~\ref{fig:example_data_pit},~\ref{fig:example_pot_pit}.

To validate the developed approach for a non-unitary $S$ matrix, I examine a simple, exactly solvable two-channel model with square-well potentials, which has been thoroughly studied by A.~N.~Kamal and H.J.Kreuzer  \cite{Kamal1970}. In their analysis, they investigate $S$-wave scattering of two spinless particles, where the first channel consists of particles with equal masses $m_1 = 1~\mathrm{GeV}$, and the second channel consists of particles with equal masses $m_2 = 1.05~\mathrm{GeV}$. The coupled
Schrodinger equations are
\begin{multline}
	\frac{d^{2}\psi_{1}}{dr^2}+(q_{1}^{2}-V_{1}(r))\psi_{1}=V_{12}\psi_{2},\\
		\frac{d^{2}\psi_{2}}{dr^2}+(q_{2}^{2}-V_{2}(r))\psi_{1}=V_{21}\psi_{1}.
	\label{model1}
\end{multline}
The momenta in the two channels are related by $q_{2}^{2} = q_{1}^{2} - q_{0}^{2}$, where $q_{0}^{2} = m_{1}E_{1} = 0.1~\mathrm{GeV}^{2}\approx 2.57~\mathrm{fm}^{-2}$  and $E_{0} = 2(m_{2} - m_{1})$ defines the inelastic threshold energy. The potential strengths are assumed to be $V_{i} = U_{i}/m_{i} = 10~\mathrm{MeV}$ for each channel, while the coupling potentials are $V_{12} = U_{12}/m_{1} \approx V_{21} = U_{21}/m_{2} = -10~\mathrm{MeV}$, with a common range of $r =a=1.973~\mathrm{fm}$.

\begin{figure}[h]
	\centerline{\includegraphics[width=0.5\textwidth]{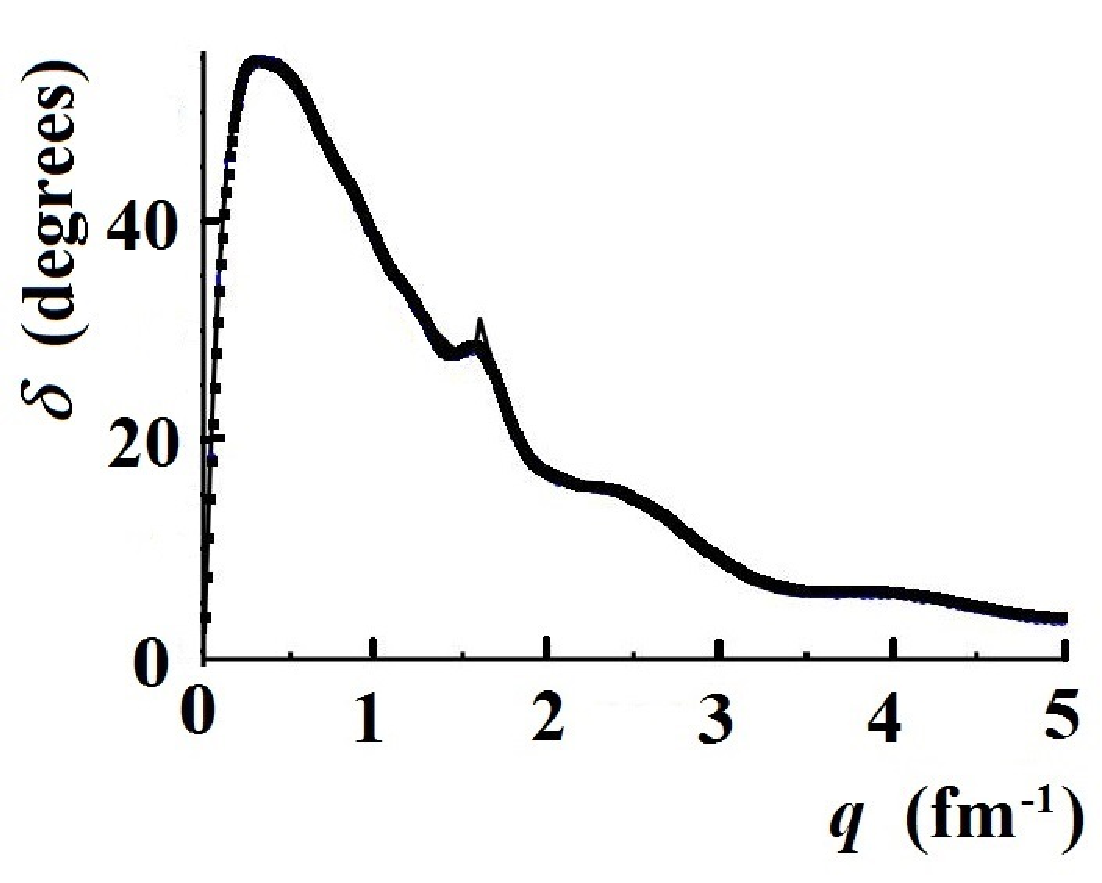}}
	\caption{\label{fig:2chPhaseOld} 
		Phase shift for the model described by Eqs.~(\ref{model1}). 
		Results from Method 1 (solid circles) are compared with the initial data (solid line).}
\end{figure}

\begin{figure}[h]
	\centerline{\includegraphics[width=0.5\textwidth]{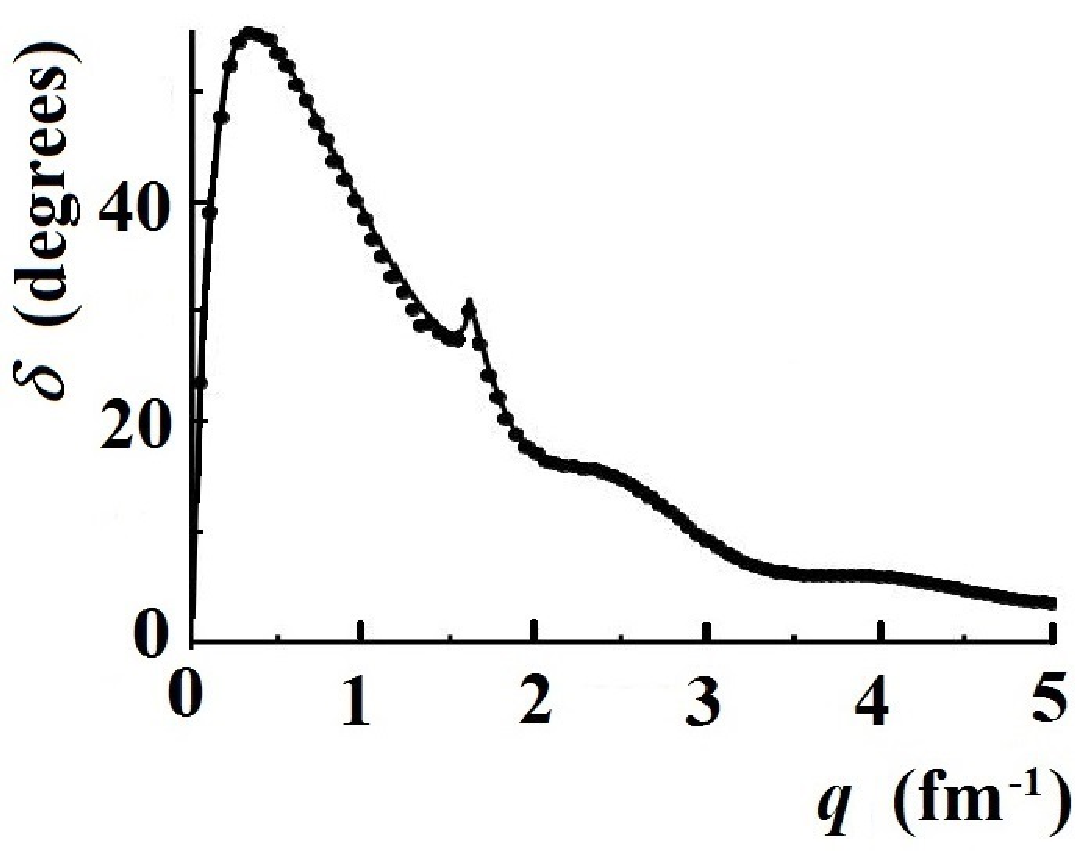}}
	\caption{\label{fig:2chPhaseNew} Phase shift for model of Eqs.~(\ref{model1}).
		Results from Method 2 (solid circles) are compared with the initial data (solid line).
	}
\end{figure}

\begin{figure}[h]
	\centerline{\includegraphics[width=0.5\textwidth]{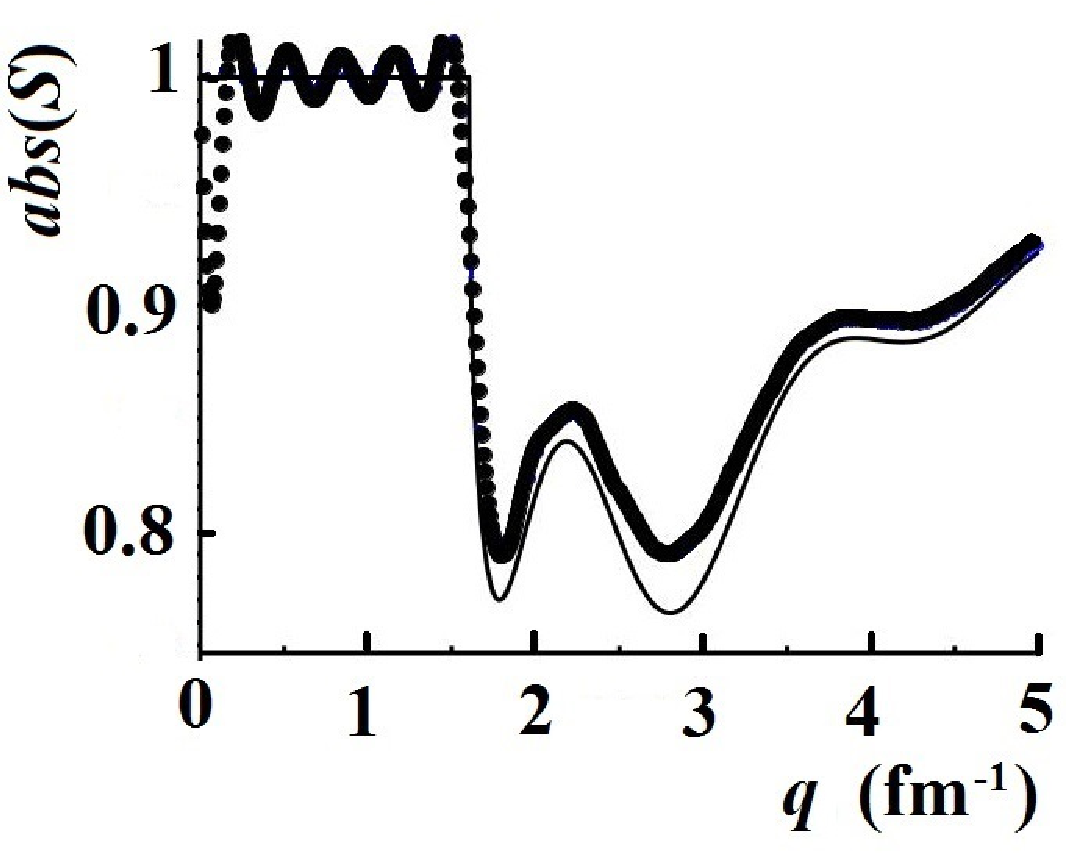}}
	\caption{\label{fig:absSold} Absolute value of $S$ matrix for model of Eqs.~(\ref{model1}).
	Results from Method 1 (solid circles) are compared with the initial data (solid line).
	}
\end{figure}

\begin{figure}[h]
	\centerline{\includegraphics[width=0.5\textwidth]{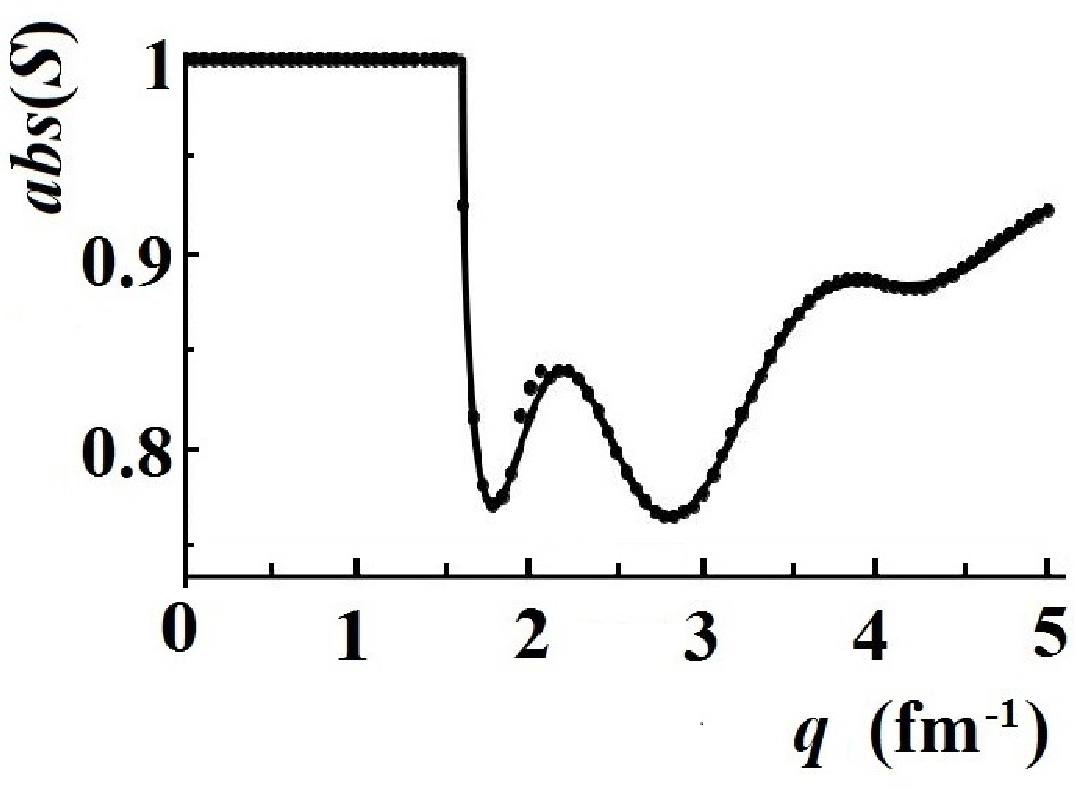}}
	\caption{\label{fig:absSnew} Absolute value of $S$ matrix for model of Eqs.~(\ref{model1}).
	Results from Method 2 (solid circles) are compared with the initial data (solid line).
	}
\end{figure}

\begin{figure}[h]
	\centerline{\includegraphics[width=0.5\textwidth]{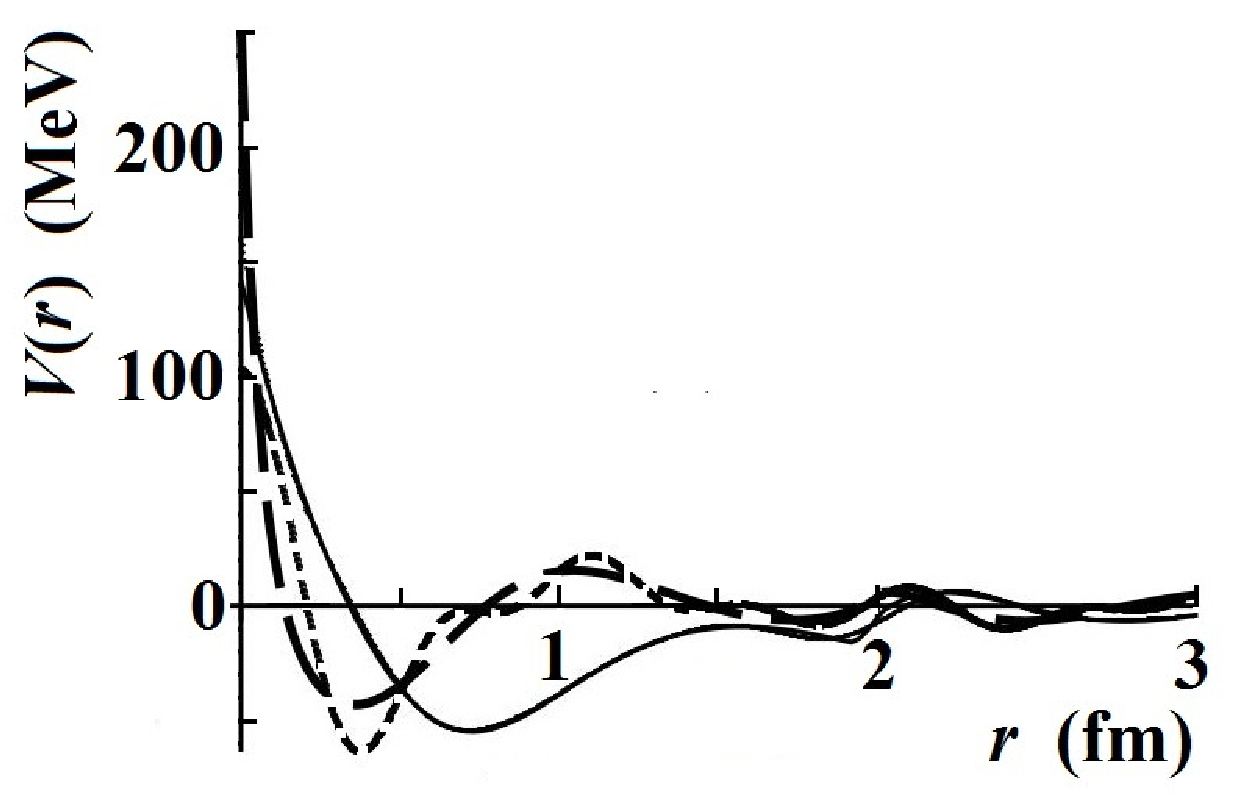}}
	\caption{\label{fig:Vmodel} Calculated optical potential for the model described by Eqs.~(\ref{model1}). The real part, $Re(V)$, is shown by solid lines (nearly overlapping for Methods 1 and 2). The imaginary part, $Im(V)$, is represented by the long dashed line for Method 1 and the short dashed line for Method 2.
	}
\end{figure}

The phase shift $\delta$ and the modulus of the $S$ matrix $|S|$ for the first channel were computed as functions of the momentum $q = q_{1}$. These results were then employed to solve the inverse problem for the first channel, assuming that the second channel is modeled by the optical potential (OP). Two methods were applied: the approach outlined in Refs.~\cite{MyAlg1,MyAlg2,MyAlg3} (Method 1), and the method developed in the present work (Method 2). 
Figs.~\ref{fig:2chPhaseOld} and  \ref{fig:2chPhaseNew} present the phase shifts for the OPs obtained using Methods 1 and 2, respectively. Although both methods provide a generally good description of the phase shifts, Method 1 smooths out the peak, whereas Method 2 captures it more satisfactorily.  
The OP obtained using Method 1 provides a significantly poorer description of the $S$-matrix modulus   compared to the OP obtained using Method 2, as illustrated in Figs.~\ref{fig:absSold} and \ref{fig:absSnew}.
Fig.~\ref{fig:Vmodel} compares the OPs obtained by both methods. The real parts of these potentials show only minor differences, whereas the discrepancy between their imaginary parts is substantial. A kink at $r \approx 2~\mathrm{fm}$ in the real part of the optical potential (OP) obtained via Method 1 indicates some sensitivity to the range of the initial potential.

\begin{figure}[h]
	\centerline{\includegraphics[width=0.5\textwidth]{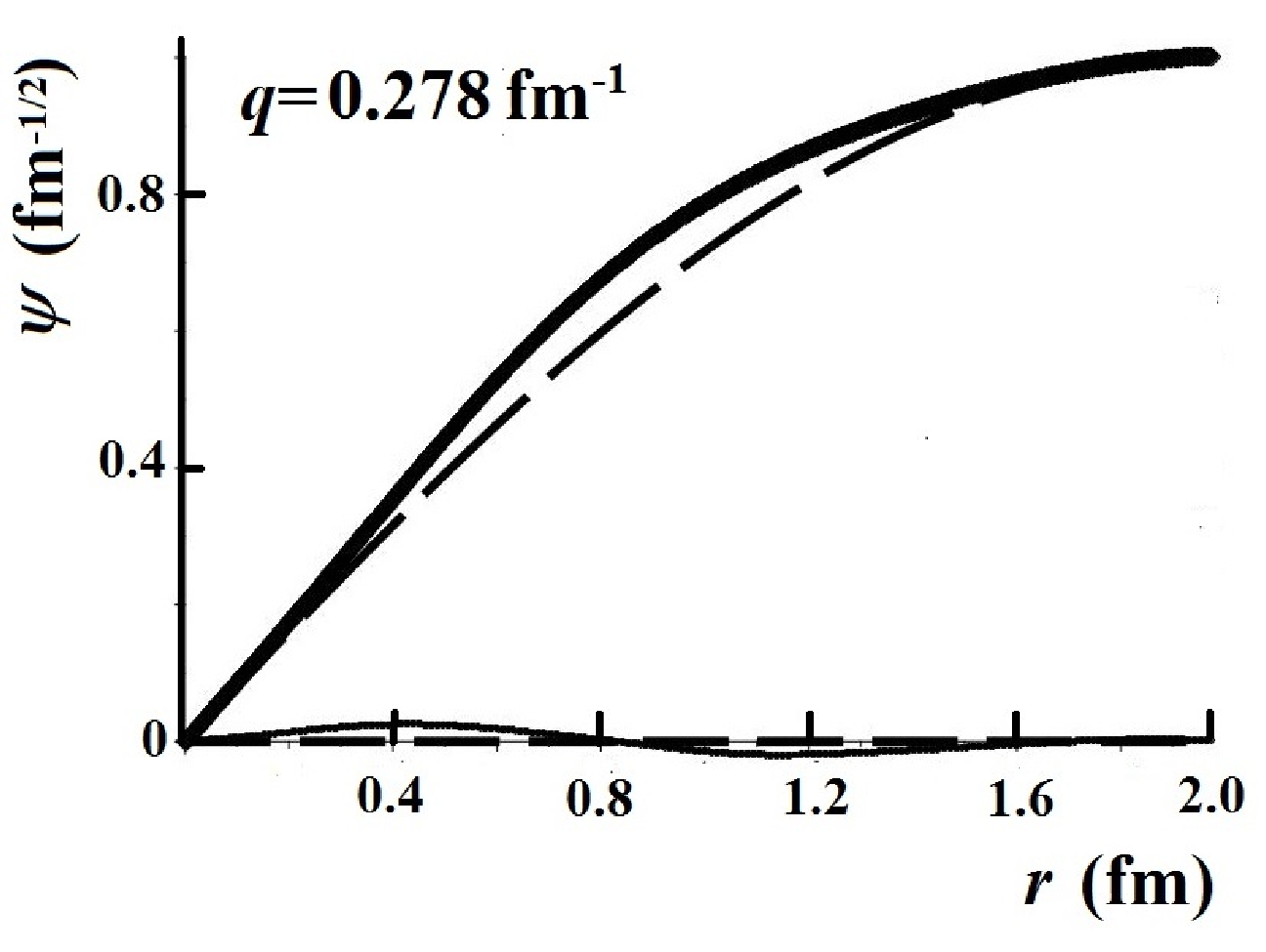}}
	\caption{\label{fig:wave1} Wave function in the first channel for the model defined by Eqs.~(\ref{model1}). Exact model: $Re(\psi)$—long-dashed line; $Im(\psi)$—thin dashed line. Calculated with the inversion potential: $Re(\psi)$—thick solid line; $Im(\psi)$—thin solid line. Momentum $q = 0.278$~fm$^{-1}$.}
\end{figure}

The wave functions calculated with the optical potential obtained by Method 2 are shown in Figs.~\ref{fig:wave1},~\ref{fig:wave1},  and \ref{fig:wave1}. The wave functions were calculated for three values of the momentum $q$ --- below, near,  and above the threshold. These wave functions are compared to those obtained from the exact model of Eqs.~\ref{model1}. At asymptotic range $r\gtrsim 2$~Fm, the OP yields wave functions nearly identical to the exact solutions. For smaller $r$, noticeable deviations appear when $q$ approaches the threshold $q_0 \approx 1.6~\mathrm{fm}^{-1}$. Below the threshold, the optical potential introduces a small imaginary component to the wave function, absent in the exact solution, but this component remains minor relative to the real part.

\begin{figure}[h]
	\centerline{\includegraphics[width=0.5\textwidth]{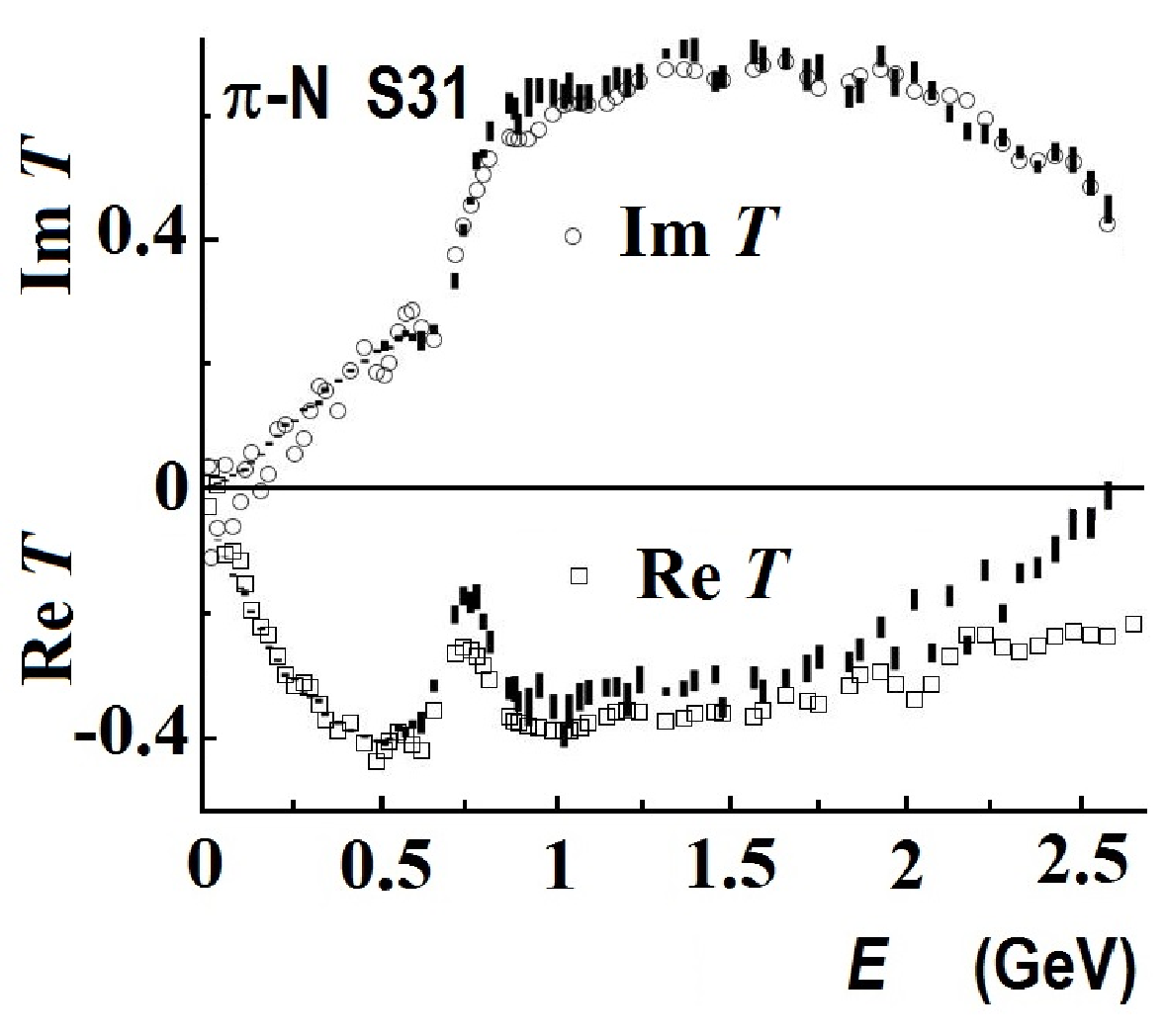}}
	\caption{\label{TS31old} 
				$S31\ \pi-N$ $T$-matrix values calculated with inversion potential obtained with Method 1 are shown by circles and squares;
		 data are from 
		\cite{TpiN_data}.
	}
\end{figure}

\begin{figure}[h]
	\centerline{\includegraphics[width=0.5\textwidth]{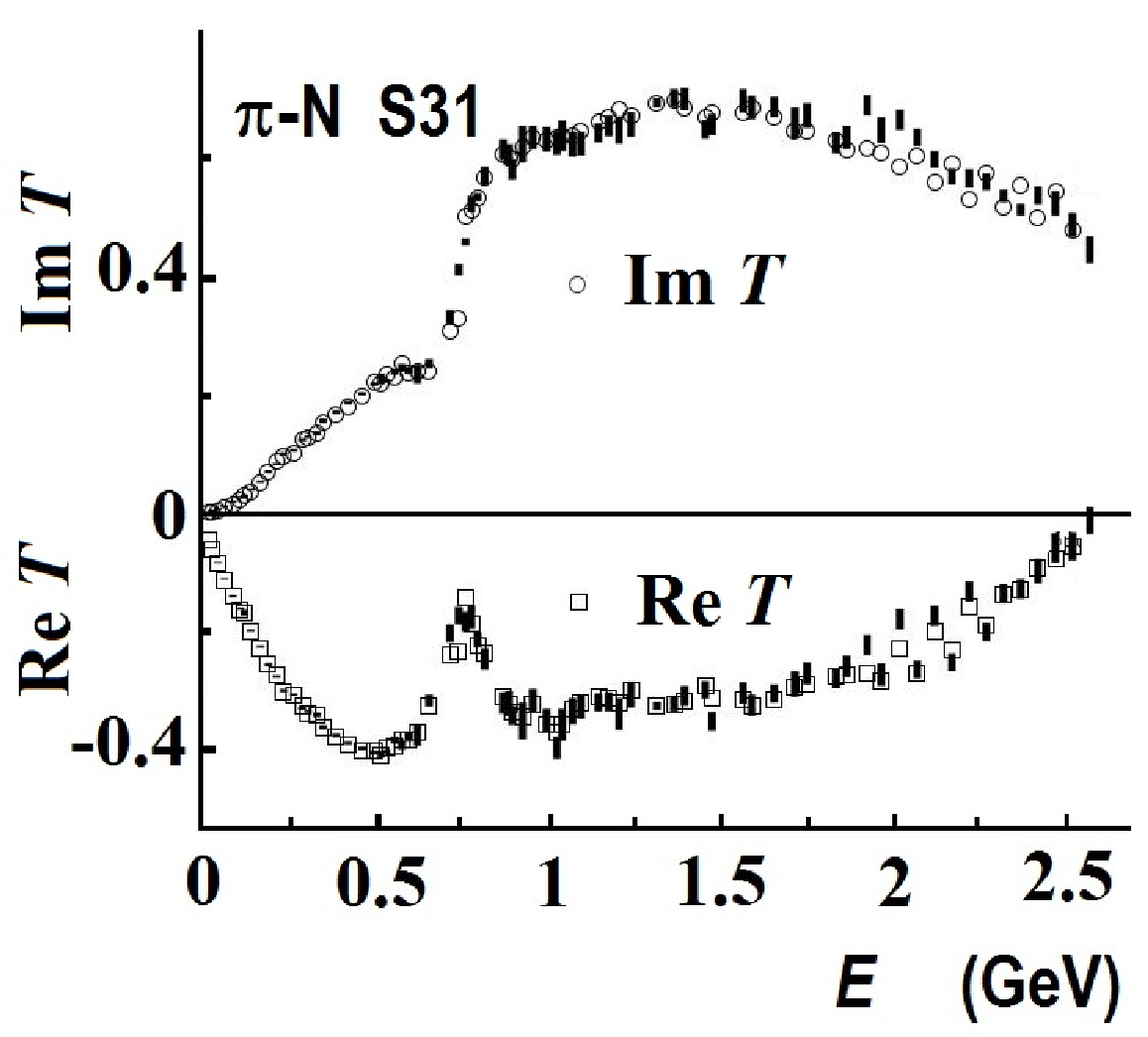}}
	\caption{\label{TS31new}  
		$S31\ \pi-N$ $T$-matrix values calculated with inversion potential obtained with Method 2 are shown by circles and squares;
		data are from 
		\cite{TpiN_data}.
		}
\end{figure}

To validate the developed methods against experimental data, the partial-wave analysis (PWA) data for the $\pi N$ $S31$ state with an inelastic resonance at $E_{\mathrm{lab}} \approx 800$~MeV were analyzed. Method 1 provides a coarse estimate of the optical potential, describing the initial data reasonably well (see Fig.\ref{TS31old}) and determining its effective range, $R$. This estimate of $R$ was subsequently used in Method 2 to derive an optical potential that fits the PWA data with accuracy exceeding the requirements for modern $\pi N$ and $NN$ scattering data (see Fig.~\ref{TS31new}).

\begin{figure}[h]
	\centerline{\includegraphics[width=0.5\textwidth]{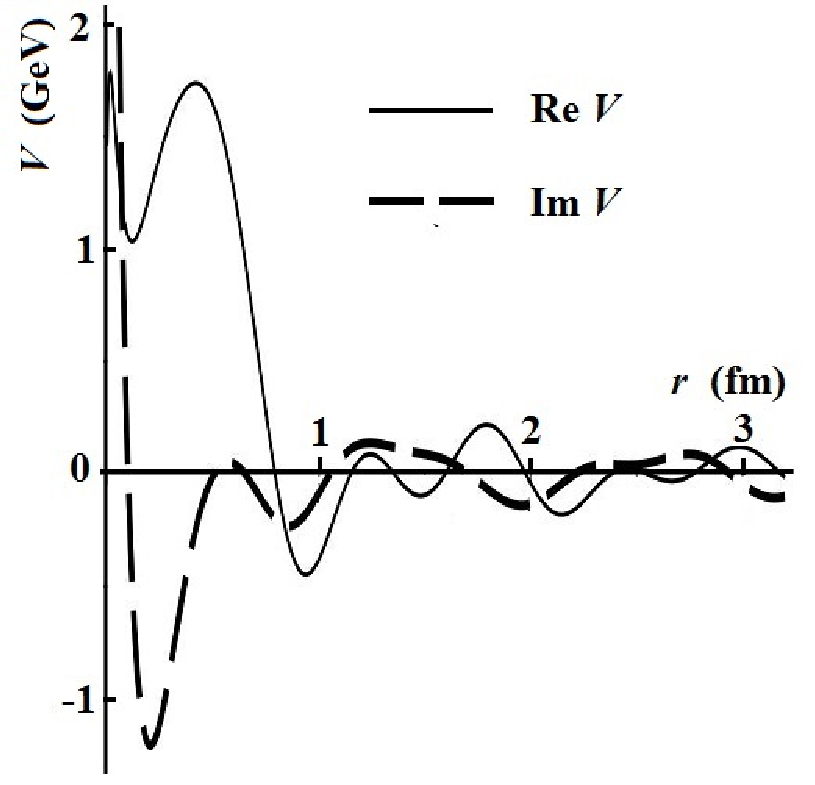}}
	\caption{\label{TS31pots} Inversion   $\pi-N$ potential obtained with Method 2 for 
		$S31$ partial wave.
	}
\end{figure}

The resulting optical potential exhibits a characteristic range of approximately 2fm and oscillates at larger distances (see Fig.\ref{TS31pots}). An accurate fit to the data in Fig.~\ref{TS31new} necessitates calculations up to $r \approx 15$~fm. Oscillations beyond $r > 2$~fm can be reduced by preprocessing input data with significant uncertainties; a method for this preprocessing is currently under development.

In summary, the developed method provides a separable kernel for the Marchenko equation (and related equations), enabling both interpolation and extrapolation for any physically reasonable dependence of the $S$ matrix (or spectral densities). This approach is readily extendable to coupled-channel cases, with a numerical implementation currently in progress for application to $NN$ and $\pi N$ scattering.

The reconstructed $S_{31}$-wave $\pi N$ complex potentials may be requested from the author in the Fortran code.

\begin{figure}[h]
	\centerline{\includegraphics[width=0.5\textwidth]{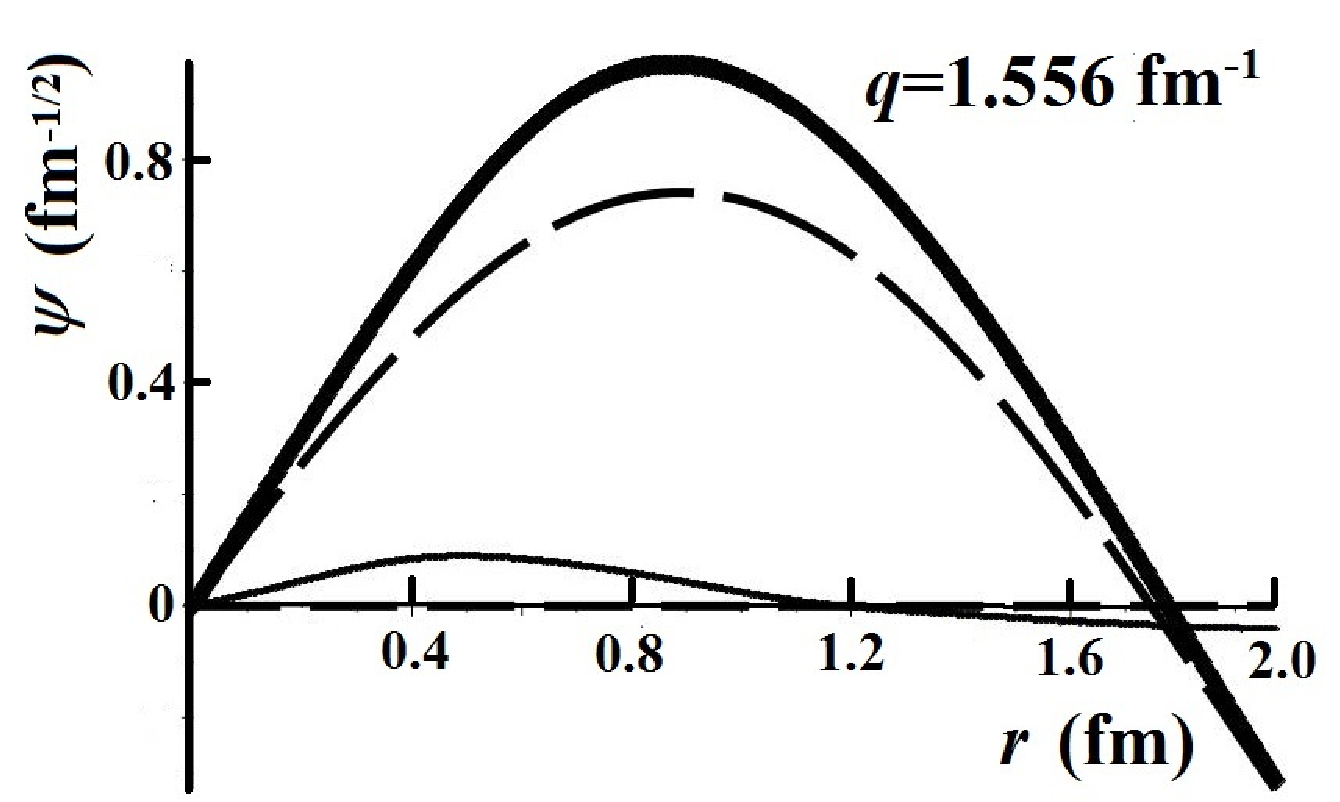}}
	\caption{\label{fig:wave2} Same as Fig.~\ref{fig:wave1}, but $q=1.556$~fm$^{-1}$.}
\end{figure}

\begin{figure}[h]
	\centerline{\includegraphics[width=0.5\textwidth]{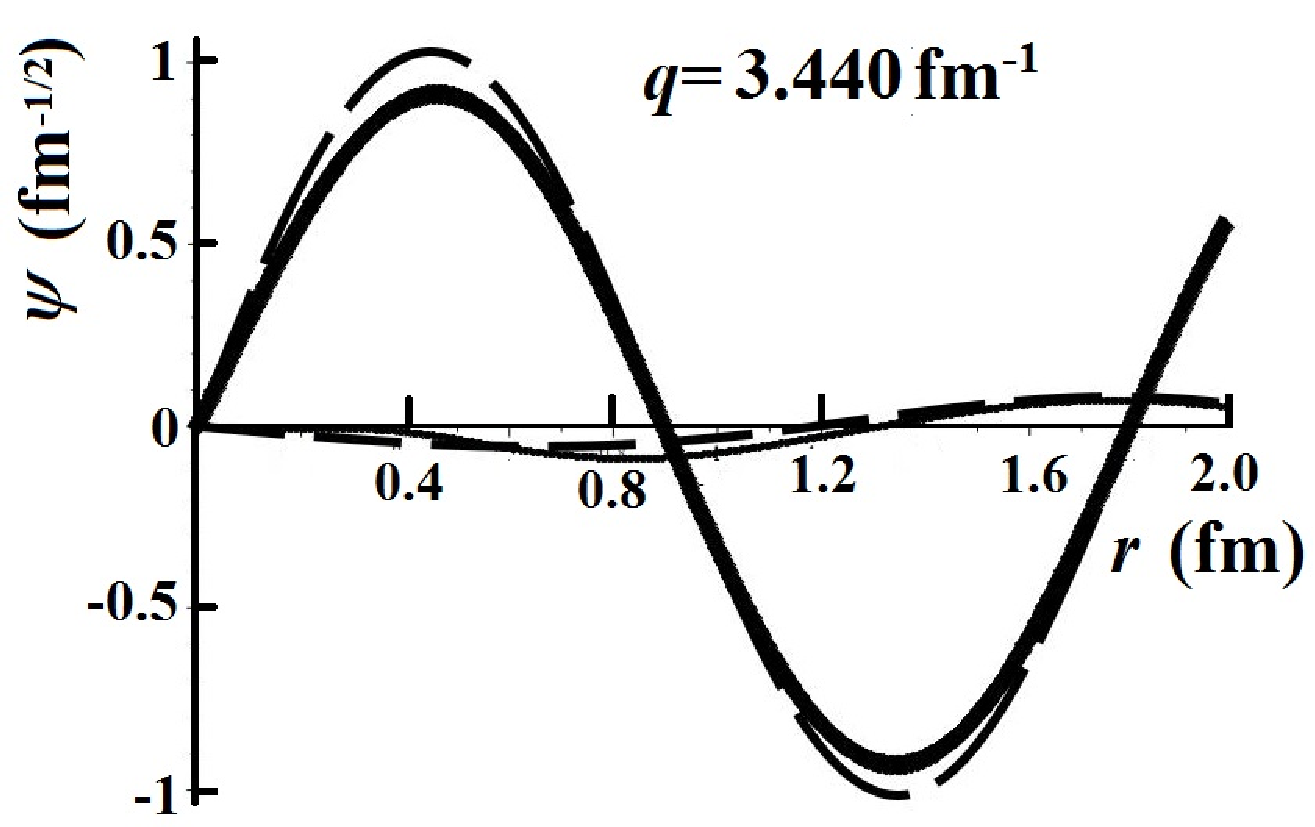}}
	\caption{\label{fig:wave3} Same as Fig.~\ref{fig:wave1}, but $q=3.440$~fm$^{-1}$.}
\end{figure}

\end{document}